\pdfoutput=1
\documentclass[sn-mathphys]{sn-jnl}
\usepackage{graphicx,amsmath,amssymb,comment,geometry,xcolor,manyfoot}
\begin{document}

\title{Non-coplanar gravitational lenses and the ``communication bridge''}

\author*[1]{\fnm{Viktor T.} \sur{Toth}}\email{vttoth@vttoth.com}

\affil*[1]{\orgaddress{\city{Ottawa}, \postcode{K1N~9H5}, \state{ON}, \country{Canada}}}

\abstract{We investigate the propagation of light signals across multiple gravitational lenses, with particular emphasis on the ``communication bridge'' scenario of two lenses with collinear source and observer. The lenses are assumed to be non-coplanar, far enough from one another for each lens to be treated independently as thin lenses in the limit of weak gravity. We analyze these scenarios using several different tools, including geometric optics, photon mapping, wave optics and ray tracing. Specifically, we use these tools to assess light amplification and image formation by a two-lens system. We then extend the ray tracing analysis to the case of multiple non-coplanar lenses, demonstrating the complexity of images that are projected even by relatively simple lens configurations. We introduce a simple simulation tool that can be used to analyze lensing by non-coplanar gravitational monopoles in the weak gravity limit, treating them as thin lenses.}

\maketitle

\section{Introduction}

In previous studies, we investigated gravitational lensing \citep{Schneider-Ehlers-Falco:1992}. Our focus was on the utility of the solar gravitational lens for image formation and its possible future use for astronomy \citep{SGL2017,SGL2019b,SGL2020a,SGL2020b,SGL2020c}, but our discussion and findings were generic, applicable to any gravitational lens, including lenses with complex mass distribution. However, to date we only studied mass distributions that can be modeled as single thin lenses in the paraxial approximation commonly used in optics.

This time, turn our attention to something different: light propagation through multiple lens planes from source to receiver. Our investigation is motivated in part by the suggestion that an advanced civilization may use multiple gravitational lenses---specifically pairs of gravitational lenses, one at each end of a communications ``bridge''---for efficient communication between stellar systems \citep{Maccone:2011}.

Our goal is to investigate these cases, in particular the two-lens ``bridge'', and derive consistent results characterizing the light amplification of two-lens and multiple-lens systems, using a variety of toolsets available. We explore geometric optics and related methods as well as the wave theory of light. The advantage of the wave theory is that it is robustly rooted in Maxwell's theory, set on the curved spacetime background of a weak gravitational field. Geometric optics, on the other hand, yields solutions that can be followed and generalized more easily. The two families of solutions must, of course, be consistent; this should be evident from the mathematics alone, notably the relationship between light ray paths in geometric optics vs. the paths that are picked by a commonly used approximation scheme for rapidly oscillating integrals, the stationary phase approximation. We demonstrate that, in particular, for the two-lens system all approaches yield consistent solutions.

This paper is organized as follows.
First, in Section~\ref{sec:geom}, we investigate the two-lens system using the tools of geometric optics. We derive a generic solution, but also address the specific case of a symmetric system, which substantially simplifies the problem and helps gaining insight into the limitations of the ``bridge'' solution.
Next, in Section~\ref{sec:photons} we re-examine the bridge by borrowing and adapting a technique from computer graphics, photon mapping (similar in concept to ``ray shooting''). Though the fundamentals are related---photon mapping is also rooted in geometric optics---the method of computation is different, and thus serves as an important verification of our main results.
We then move on to wave optics in Section~\ref{sec:waves}. Analyzing the complex amplitude of incoming light as it passes through two lenses, we obtain a result that is consistent with the results derived from geometric methods.
Finally, we explore ray tracing in Section~\ref{sec:raytrace}. The main disadvantage of conventional ray tracing is that it only offers qualitative results, it does not allow us to accurately estimate light amplification. On the other hand, ray tracing can be easily extended to model complex scenarios. We demonstrate how even a two-lens system can produce surprisingly complex images, and show some results with four lenses.
We present our conclusions in Section~\ref{sec:disc}.

\section{Geometric optics}
\label{sec:geom}

Gravitational lenses are remarkably powerful, courtesy of their very large size, which yields substantial light amplification and angular resolution, far in excess of any manufactured apparatus. At the same time, these lenses are characterized by severe spherical aberration and, in the case of actual astrophysical objects, by often significant deviations from spherical or axial symmetry \citep{SGL2021a}.

\begin{figure*}[t]
\begin{center}
\includegraphics{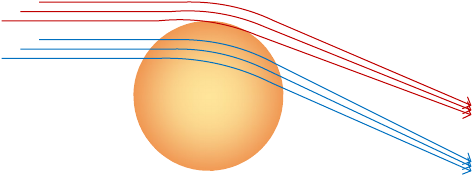}
\end{center}
\caption{\label{fig:deflection}Deflection of light rays by a monopole gravitational lens. Light rays with the same impact parameter converge azimuthally. Light rays with different impact parameters
diverge radially.}
\end{figure*}

Even a ``perfect'' gravitational lens, i.e., a gravitational monopole, leaves a lot to be desired as a lens. As shown in Fig.~\ref{fig:deflection}, light rays in the vicinity of a monopole lens converge azimuthally, but diverge radially \citep{vonEshleman:1979}, as light rays with greater impact parameters are deflected by a smaller angle in the weaker gravitational field. The result is a lens that suffers from significant negative spherical aberration, reducing its ability to focus light.

Having explored in-depth lensing and image formation by compact gravitational lenses characterized by a monopole potential with multipole perturbations, we now turn our attention to a different scenario: light that is deflected by not one but several gravitational lenses in succession. A particular case of interest is that of the gravitational lens ``bridge'': the notion that a pair of gravitational lenses, one near the source and one near the receiver, may yield extraordinary amplification, and thus may be used by a sufficiently advanced technological civilization for effective interstellar communication \citep{Maccone:2011}.

\begin{figure*}[t]
\begin{center}
\includegraphics{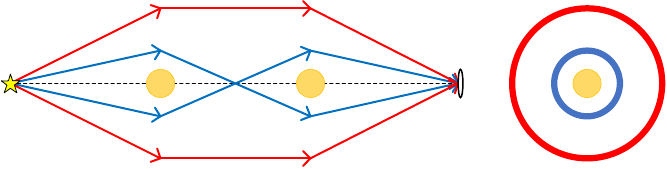}
\end{center}
\caption{\label{fig:tworings}A generic two-lens system, 
viewed from the optical axis in the focal region of the second lens, produces two concentric Einstein rings unless the light rays are blocked by the 
lenses
themselves.}
\end{figure*}

We begin with a simple qualitative assessment of the axisymmetric two-lens system. Depicted in Fig.~\ref{fig:tworings}, we can see that when the source, the two lenses, and the receiver are colinear, there are two distinct light paths from source to receiver: a ``straight'' path that involves light rays passing both lenses on the same side, and a ``crossover'' path that is characterized by smaller impact parameters and greater deflection angles.

For an observer at the receiving end, equipped with an instrument of sufficient angular resolution, this yields an image consisting of two concentric Einstein rings that appear around the second lens. However, we must note that in astrophysical scenarios such as those contemplated by \citep{Maccone:2011}, the radii of the two rings will be very similar, so it may not be possible for an observer to resolve the two rings as distinct. Nonetheless, when assessing the optical characteristics of the lens, we must keep this in mind that the observer receives light from the combined light collecting area of the two concentric rings.

\begin{figure}
\begin{center}
\includegraphics[width=\textwidth]{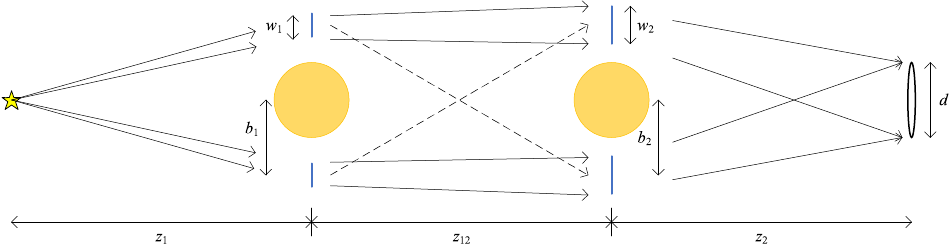}
\end{center}
\caption{\label{fig:bridge}The geometry of the communication ``bridge''.}
\end{figure}

Moving beyond qualitative statements, we now turn our attention to a geometric optics analysis of the two-ring system, using tools developed in \cite{SGL2023e}. Let us consider a parameterization of the two-lens ``bridge'' as shown in Fig.~\ref{fig:bridge}.

As noted, there are really two impact parameters at the second lens, corresponding to the two spots of light that appear near the first lens, one on the same side as the vantage point at the second lens, one on the opposite side (represented by dashed lines in Fig.~\ref{fig:bridge}). Therefore, the observing telescope sees two Einstein rings around the second lens, with slightly different impact parameters (and corresponding angular radii) formed from light coming from the same side vs. the opposite side of the first lens.

\begin{figure}
\includegraphics[scale=0.75]{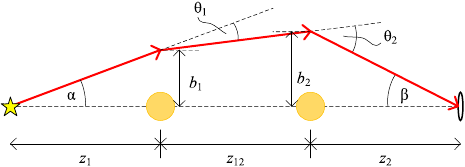}\hskip 0.5in\includegraphics[scale=0.75]{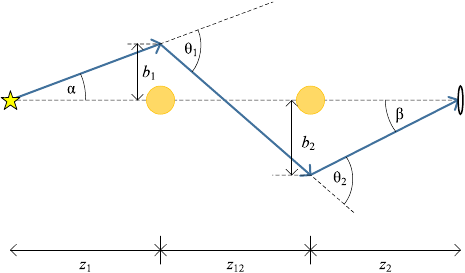}
\caption{\label{fig:bgeom}The geometry of the bridge for the ``straight'' and ``crossover'' cases.}
\end{figure}

It is not hard in principle to calculate the geometry of these cases but unfortunately the analytical results, being the solutions of quartic equations, are not very practical. Still, it is instructive to see how these solutions emerge. Simply reading Fig.~\ref{fig:bgeom} (left), we can establish, for the ``straight'' case, the following system of equations:
\begin{align}
\alpha+\beta&{}=\theta_1+\theta_2,\\
\alpha&{}=\frac{b_1}{z_1},\\
\beta &{}=\frac{b_2}{z_2},\\
\frac{b_2-b_1}{z_{12}}&{}=\alpha-\theta_1,
\end{align}
and, of course, $\theta_i=2r_{gi}/b_i$ ($i=1,2$) where $r_{gi}$ is the corresponding Schwarzschild radius. Trivial algebra leads allows us to eliminate most variables with ease, leaving us with these two equations for $b_1$ and $b_2$:
\begin{align}
\frac{b_1}{z_1} + \frac{b_2}{z_2}&{}=\frac{2r_{g1}}{b_1} + \frac{2r_{g2}}{b_2},\\
\frac{b_2-b_1}{z_{12}} &{} = \frac{b_1}{z_1} - \frac{2r_{g1}}{b_1}.
\end{align}
As mentioned, this yields an impractical quartic solution. However, we can evaluate this system in the symmetric case, when the two stars are identical with identical impact parameters, that is, $r_{g1}=r_{g2}=r_g$, $z_1=z_2=z$ and $b_1=b_2=b_\parallel$:
\begin{align}
b_\parallel = \sqrt{2r_gz},
\end{align}
a familiar result.

The equations for the ``crossover'' case, shown in Fig.~\ref{fig:bgeom} (right), are similar:
\begin{align}
\alpha-\beta&{}=\theta_1-\theta_2,\\
\alpha&{}=\frac{b_1}{z_1},\\
\beta &{}=\frac{b_2}{z_2},\\
\frac{b_1+b_2}{z_{12}}&{}=\theta_1-\alpha,
\end{align}
which simplify to
\begin{align}
\frac{b_1}{z_1}-\frac{b_2}{z_2}&{}=\frac{2r_{g1}}{b_1}-\frac{2r_{g2}}{b_2},\\
\frac{b_1+b_2}{z_{12}}&{}=\frac{2r_{g1}}{b_1}-\frac{b_1}{z_1}.
\end{align}
As before, the general solution of this system is a quartic solution that has little practical utility. In the case of the symmetric bridge, however, we can establish $b_1=b_2=b_\times$, $r_{g1}=r_{g2}=r_g$ and $z_1=z_2=z$, which yields
\begin{align}
\frac{2b_\times}{z_{12}}=\frac{2r_g}{b_\times}-\frac{b_\times}{z},
\end{align}
or
\begin{align}
b_\times=\sqrt{\frac{2r_gzz_{12}}{2z+z_{12}}}\simeq\sqrt{2r_gz}\left(1 - \frac{z}{z_{12}}\right)=b_\parallel\left(1 - \frac{z}{z_{12}}\right).
\end{align}

We thus established that a symmetric communications ``bridge'' collects light at the first star through the light collecting area that corresponds to two concentric Einstein rings of nearly equal radius. To compute the amplification of the ``bridge'', however, we must also be able to estimate the width of these Einstein ring annuli.

To do so, we again remind ourselves that gravitational lenses disperse light: initially parallel rays of light diverge, and even a small difference in impact parameter (corresponding to the diameter of the observing telescope's aperture) can have a significant impact on the observed width of the Einstein ring annulus.

Assuming that divergent light rays are deflected by the same angle at each lens would lead us to the naive result that, if the distance between two neighboring light rays is $w_1$ at the first lens, it will be $w_2=(z_1+z_{12})w_1/z_1$ at the second lens, and $w=(z_1 + z_{12} + z_2)w_1/z_1$ at the image plane.

In actuality, if a light ray with impact parameter $b_1$ is deflected by the angle $\theta_1=2r_{g1}/b_1$, another ray with impact parameter $b_1+w_1$ will be deflected by
\begin{align}
\theta_1'=\frac{2r_{g1}}{b_1}\left(1-\frac{w_1}{b_1}\right).
\end{align}
Consequently, at the second lens the distance between the two light rays will increase to
\begin{align}
w_2=(z_1+z_{12})\frac{w_1}{z_1} + z_{12}\frac{2r_{g1}w_1}{b_1^2}.
\end{align}

Once again, at the second lens if a light ray with impact parameter $b_2$ is deflected by $\theta_2=2r_{g2}/b_2$, a light ray with impact parameter $b_2+w_2$ will be deflected by
\begin{align}
\theta_2'=\frac{2r_{g2}}{b_2}\left(1-\frac{w_2}{b_2}\right).
\end{align}

Consequently, the projected distance between the light rays as they intersect the image plane will be given as
\begin{align}
w&{}=(z_1+z_{12}+z_2)\frac{w_1}{z_1}+(\theta_1-\theta_1')(z_{12}+z_2)+(\theta_2-\theta_2')z_2\nonumber\\
&{}=(z_1+z_{12}+z_2)\frac{w_1}{z_1}+(z_{12}+z_2)\frac{2r_{g1}w_1}{b_1^2}+z_2\frac{2r_{g2}w_2}{b_2^2}\nonumber\\
&{}=\left[\frac{z_1+z_{12}+z_2}{z_1}+\frac{2r_{g1}(z_{12}+z_2)}{b_1^2}+\frac{2r_{g2}z_2}{b_2^2}\left(\frac{z_1+z_{12}}{z_1}+\frac{2r_{g1}z_{12}}{b_1^2}\right)
\right]w_1.
\end{align}

If we equate $w$ with the aperture of the observing telescope in the image plane, $w=d$, we can use this result to compute the width of the annulus around the first gravitational lens that represents the light collecting area. (Though tempting to call it an Einstein ring, this is not a directly visible Einstein ring.) Specifically in the case of the symmetric bridge, $r_{g1}=r_{g2}=r_g$, $z_2=z_1=z$, we have two impact parameters, for the `straight' and `crossover' cases:
\begin{align}
b_1=b_2=b_\parallel&{}=\sqrt{2r_g z}\\
b_1=b_2=b_\times&{}=\sqrt{\frac{2r_gzz_{12}}{2z+z_{12}}}.
\end{align}
Carrying out the substitution, we obtain, for the two cases,
\begin{align}
d&{}=\left[\frac{2z+z_{12}}{z}+\frac{2r_{g}(z+z_{12})}{b_\parallel^2}+\frac{2r_gz}{b_\parallel^2}\left(\frac{z+z_{12}}{z}+\frac{2r_{g}z_{12}}{b_\parallel^2}\right)
\right]w_\parallel=4\frac{z+z_{12}}{z}w_\parallel\\
d&{}=\left[\frac{2z+z_{12}}{z}+\frac{2r_{g}(z+z_{12})}{b_\times^2}+\frac{2r_gz}{b_\times^2}\left(\frac{z+z_{12}}{z}+\frac{2r_{g}z_{12}}{b_\times^2}\right)
\right]w_\times 
=4\frac{(2z+z_{12})(z+z_{12})}{zz_{12}}w_\times,
\end{align}
where we used $w_1=w_\parallel$ and $w_1=w_\times$ to represent the width of the annulus for the ``straight'' and ``crossover'' cases. In the limit $z_{12}\gg z$, both expressions simplify to
\begin{align}
w_1=\frac{z}{4z_{12}}d.
\end{align}
The light collecting area corresponding to this annulus with radius $b$ is given by $2\pi w_1 b$. Two such concentric annuli will have the light collecting area $4\pi w_1 b=\pi bdz/z_{12}$ at distance $z$ from the source, corresponding to the solid angle
\begin{align}
\Theta_{\tt 2GL}=\frac{\pi bdz}{z^2z_{12}}.
\end{align}
Comparing against the solid angle subtended by an aperture $d$ at $z_{12}$, $\Theta_d=\tfrac{1}{4}\pi d^2/z_{12}^2$, we obtain the light amplification factor for the symmetric bridge in the $z_{12}\gg z$ limit:
\begin{align}
\mu_{\tt 2GL}=\frac{\Theta_{\tt 2GL}}{\Theta_d}=\frac{\pi bdz}{z^2z_{12}}\frac{4z_{12}^2}{\pi d^2}=\frac{z_{12}}{z}\frac{4b}{d},
\end{align}
which is identical to the light amplification factor of a single gravitational lens at $z$ from the source. Therefore, the ``bridge'' configuration does not offer additional amplification.

\section{Photon mapping}
\label{sec:photons}

There is a closely related yet distinct method of estimating the light amplification of the lens, inspired by a commonly used family of computer graphics algorithms, photon mapping. In the case of computer graphics, ray tracing---tracing light rays from an observing position back towards light sources---may be supplemented in a two pass process by tracing light from the source, in order to overcome a known shortcoming of ray tracing algorithms concerning the accuracy of estimating the level of illumination.

We can not only adopt this technique to the case of the two-lens configuration, as it turns out in the axisymmetric case we can offer an analytic solution.

Again referencing Fig.~\ref{fig:bridge}, we can trace a light ray that leaves the source at angle $\alpha$. When this light ray hits the first lens plane, it is deflected; in the second lens plane, it is deflected again. Eventually, it intersects the image plane. We can thus ask a simple question: for what values of $\alpha$ will the light ray intersect the image plane within $|\rho|\le \tfrac{1}{2}d$ from the optical axis, i.e., within the aperture of an observing lens of diameter $d$?

As the system is axially symmetric throughout, the calculation turns out to be elementary, especially when we maintain the paraxial approximation. First, let us consider the case of a single lens, i.e., when $r_{g2}=0$, hence $\theta_2=0$ and the second lens contributes nothing. To depict this scenario, we use the notation $z=z_1$, $\bar{z}=z_{12}+z_2$, $b=b_1$ and $r_{g1}=r_g$. A light ray that departs the source at angle $\alpha$ will arrive at the first (and only) lens plane with an impact parameter $b=z\alpha$. It is then deflected by $\theta=2r_{g}/b=2r_{g}/z\alpha$ towards the optical axis, to arrive at the image plane at $\rho=(z+\bar{z})\alpha-\bar{z}\theta=(z+\bar{z})\alpha-2r_g\bar{z}/z\alpha$. The criterion $|\rho|\le \tfrac{1}{2}d$ can therefore be expressed as
\begin{align}
-\tfrac{1}{2}d\le(z+\bar{z})\alpha-\frac{2r_g\bar{z}}{z\alpha}\le\tfrac{1}{2}d.
\end{align}
Solving for equality and keeping only positive values of $\alpha$ (negative values just correspond to the same angle mirrored on the opposite side of the lens), we obtain
\begin{align}
\alpha_\pm=\sqrt{\frac{d^2}{16(z+\bar{z})^2}+\frac{2r_g\bar{z}}{z(z+\bar{z})}}\pm\frac{d}{4(z+\bar{z})}.
\end{align}
Considering the usual scenario of $z\ll\bar{z}$ and $d\ll r_g$, this expression simplifies to the approximate value of
\begin{align}
\alpha_\pm\simeq\sqrt{\frac{2r_g}{z}}\pm\frac{d}{4\bar{z}},\label{eq:alphapm-1GL}
\end{align}
i.e., a thin annulus characterized by the standard Einstein deflection angle $\theta=\sqrt{2r_g/z}=b/z=2r_g/b$. The corresponding solid angle is given by
\begin{align}
\Theta_{\tt 1GL}=\pi\frac{bd}{z\bar{z}}.
\end{align}
In comparison with a thin lens telescope of aperture $d$ located at $\bar{z}$, subtending the solid angle $\Theta_d=\tfrac{1}{4}\pi d^2/\bar{z}^2$, this yields the expected light amplification factor
\begin{align}
\mu_{\tt 1GL}=\frac{\Theta_{\tt 1GL}}{\Theta_d}=\frac{\pi bd}{z\bar{z}}\frac{4\bar{z}^2}{\pi d^2}=\frac{\bar{z}}{z}\frac{4b}{d},
\end{align}
a previously known result.

The same procedure can be followed when a second lens is present. In this case, we have
\begin{align}
b_1&{}=\alpha z_1,\\
\theta_1&{}=\frac{2r_{g1}}{b_1},\\
b_2&{}=\alpha(z_1+z_{12})-\theta_1z_{12},\\
\theta_2&{}=\frac{2r_{g2}}{b_2},\\
\rho&{}=\alpha(z_1+z_{12}+z_2)-\theta_1(z_{12}+z_2)-\theta_2z_2\nonumber\\
&{}=\alpha(z_1+z_{12}+z_2)-\frac{2r_{g1}(z_{12}+z_2)}{\alpha z_1}-\frac{2r_{g2}z_2}{\alpha(z_1+z_{12})-\displaystyle\frac{2r_{g1}z_{12}}{\alpha z_1}}.
\end{align}
Once again, we seek solutions that correspond to the inequality, $|\rho|\le\tfrac{1}{2}d$. Solving for equality, however, now entails solving a quartic equation for $\alpha$. Algebraic solutions exist, but the complexity of the expressions renders them impractical. Let us instead again examine the limit of the symmetric case, $z_{12}\gg z_1=z_2=z$, $r_{g1}=r_{g2}=r_g$, $b_1=b_2=b$. In this case, we have
\begin{align}
\rho\simeq z_{12}\alpha-\frac{2r_{g}z_{12}}{z\alpha },
\end{align}
and the inequality $|\rho|\le\tfrac{1}{2}d$ is solved by
\begin{align}
\alpha_\pm\simeq\sqrt{\frac{2r_g}{z}}\pm\frac{d}{4z_{12}},
\end{align}
which is identical to the single-lens case (\ref{eq:alphapm-1GL}) we looked at earlier, thus $\mu_{\tt 2GL}=\mu_{\tt 1GL}$.

Similarly, when we evaluate the generic two-lens case numerically without resorting to approximations, we recover the same result: the second lens of the ``bridge'' does not significantly contribute to light amplification in the limit of geometric optics.


~\par

\section{The wave theory}
\label{sec:waves}

In earlier work \citep{SGL2017}, we developed a comprehensive wave-theoretical description of gravitational lenses. In particular, we found that considering an incident plane wave, the dimensionless amplitude of the electromagnetic field in the image plane of a monopole gravitational lens at distance $z$ from the lens, characterized by the polar radial coordinate $\rho$ in that plane, can be characterized up to a constant phase by the Bessel function,
\begin{align}
{\cal A}=\sqrt{2\pi kr_g}
J_0\left(k\sqrt{\frac{2r_g}{z}}\rho\right).
\end{align}

\begin{figure*}[t]
\begin{center}
\includegraphics{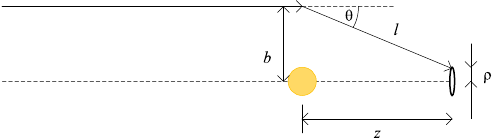}
\end{center}
\caption{\label{fig:onelens-planewave}The geometry of a plane wave encountering a gravitational lens.}
\end{figure*}

It is important to recognize the geometric significance of the term multiplying the wavenumber $k$ in the argument of the Bessel function, $\sqrt{2r_g/z}\rho=\theta\rho=b_0\rho/z$, where we used $b_0$ to denote the impact parameter corresponding to a light ray that hits the optical axis at exactly the image plane. This quantity directly characterizes the change in the length of light paths as a function of $\rho$ in the paraxial approximation. To see this, consider the following, using the notation in Fig.~\ref{fig:onelens-planewave}. First,
\begin{align}
\theta=\frac{b-\rho}{z}=\frac{2r_g}{b},
\end{align}
from which we obtain
\begin{align}
b_{\pm}=\tfrac{1}{2}\rho\pm\sqrt{\tfrac{1}{4}\rho^2+2r_gz}\simeq \pm b_0+\tfrac{1}{2}\rho.
\end{align}
The two (positive and negative) solutions $b_\pm$ correspond to two light rays passing on opposite sides of the lens. The corresponding Euclidean light paths are given by (note that the actual propagation delay also includes a logarithmic Shapiro term that, however, cancels out at our level of approximation when we calculate differences in phase):
\begin{align}
l_\pm=\sqrt{z^2+(b_0\pm\tfrac{1}{2}\rho)^2}\simeq \tilde{l}\pm\frac{1}{2}\frac{b_0}{z}\rho,
\end{align}
with $\tilde{l}=\sqrt{b_0^2+z^2}$. The difference between the two, which determines the interference pattern in the image plane, is
\begin{align}
l_+-l_-=\frac{b}{z}\rho=\sqrt{\frac{2r_g}{z}}\rho,
\end{align}
which, multiplied by $k$, is the argument of the Bessel function in the expression for ${\cal A}$.

\begin{figure*}[t]
\begin{center}
\includegraphics{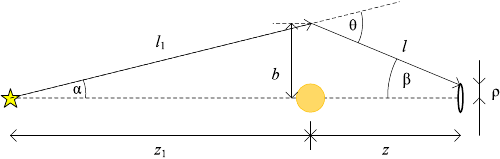}
\end{center}
\caption{\label{fig:onelens-finite}The geometry of light from a finite but large distance encountering a gravitational lens.}
\end{figure*}

Recognizing this informs us as to how we can readily adopt this formalism to the case when the incident wave comes from a light source at a finite but large distance (so the paraxial approximation still applies, and we are still allowed to treat individual ``light rays'' using the plane wave approximation), as depicted in Fig.~\ref{fig:onelens-finite}. In this case, we can write
\begin{align}
l_1&{}=\sqrt{z_1^2+b^2},\\
l&{}=\sqrt{z^2+(b-\rho)^2},
\end{align}
while
\begin{align}
\theta=\frac{2r_g}{b}=\alpha+\beta=\frac{b}{z_1}+\frac{b-\rho}{z},
\end{align}
or
\begin{align}
b^2-\frac{z_1\rho}{z+z_1}-\frac{2r_gzz_1}{z+z_1}{}&=0,
\end{align}\begin{align}
b_\pm&{}=\frac{z_1}{z+z_1}\left[\tfrac{1}{2}\rho\pm\sqrt{\tfrac{1}{4}\rho^2+\frac{z+z_1}{z_1}2r_gz}\right]\simeq
\pm
b_0
+\frac{z_1}{2(z+z_1)}\rho
=
\pm b_0+\tfrac{1}{2}\tilde\rho,
\end{align}
with $
b_0
=\sqrt{2r_gzz_1/(z+z_1)}$, $\tilde{l}_1=\sqrt{b_0^2+z_1^2}$ and $\tilde\rho=z_1\rho/(z+z_1)$. (Note that the last approximation is valid only when $\rho\ll 4b\sqrt{(z+z_1)/z_1}$, which may not hold in all cases of interest). Correspondingly, we get
\begin{align}
l_{1\pm}&{}=\sqrt{z_1^2+(b_0\pm\tfrac{1}{2}\tilde\rho)^2}\simeq\tilde{l}_1\pm\frac{1}{2}\frac{b_0}{z+z_1}\rho,\\
l_\pm&{}=\sqrt{z^2+(b_0\pm\tfrac{1}{2}\tilde\rho)^2}\simeq\tilde{l}\pm\frac{1}{2}\frac{z_1b_0}{z(z+z_1)}\rho,
\end{align}
and the change in Euclidean path length is characterized by
\begin{align}
l_{1+}+l_+-l_{1-}-l_-=\frac{b_0}{z}\rho=\sqrt{\frac{2r_gz_1}{z(z+z_1)}}\rho.
\end{align}


This, then, must be the argument of $J_0$ characterizing the difference in light paths, i.e., governing constructive and destructive interference, which allows us to extend the wave-optical formalism to the case when the source is at a finite distance from a single lens:
\begin{align}
{\cal A}_{\tt 1GL}=\sqrt{2\pi kr_g}J_0\left(k\sqrt{\frac{2r_gz_1}{z(z+z_1)}}\rho\right).
\end{align}
The corresponding PSF is given as the square of this dimensionless amplitude, i.e.,
\begin{align}
{\tt PSF}_{\tt 1GL}=2\pi kr_g J_0^2\left(k\sqrt{\frac{2r_gz_1}{z(z+z_1)}}\rho\right).
\label{eq:PSF1}
\end{align}
The squared Bessel function in this PSF oscillates, with an average amplitude of $\left<J_0^2(x)\right>\simeq 1/\pi x$, allowing us to write,
\begin{align}
\left<{\tt PSF}_{\tt 1GL}\right>=\frac{2 r_g}{\rho} \sqrt{\frac{z(z+z_1)}{2r_gz_1}}\simeq \frac{\theta z}{\rho}, 
\label{eq:PSF1avg}
\end{align}
where the final approximate form is given assuming $z\gg z_1$, and also using $\theta=\sqrt{2r_g/z_1}$. 

In the case of the two-lens system, the image plane of the first lens corresponds to the lens plane of the second lens, which implies the substitutions $r_g=r_{g1}$, $z=z_{12}$, $\rho=b_2$. This allows us to form, for the amplitude of the system, an integral in the second lens plane that convolves the amplitude of the first lens with the properties of the second:
\begin{align}
{\cal A}_{\tt 2GL}(\rho)&{}=
\frac{k}{\tilde{z}_2}\int_0^\infty db_2~b_2
{\cal A}_{\tt 1GL}(b_2)J_0\left(k\frac{b_2}{\tilde{z}_2}\rho\right)\exp\left[ik\left(\frac{b_2^2}{2\tilde{z}_2}-2r_{g2}\ln kb_2\right)\right]\nonumber\\
&{}=
\sqrt{2\pi kr_{g1}}
\frac{k}{\tilde{z}_2}\int_0^\infty db_2~b_2J_0\left(k\sqrt{\frac{2r_{g1}z_1}{z_{12}(z_{12}+z_1)}}b_2\right)J_0\left(k\frac{b_2}{\tilde{z}_2}\rho\right)\nonumber\\
&\hskip 1.5in{}\times\exp\left[ik\left(\frac{b_2^2}{2\tilde{z}_2}-2r_{g2}\ln kb_2\right)\right].
\end{align}
For realistic parameter values, the first of these two Bessel functions oscillates far more rapidly than the second, allowing us to target it using the method of stationary phase. To do so, we first replace this occurrence of $J_0$ with its approximation for large arguments, which we are allowed to do since $b_2\gtrsim R_\star$ for physically relevant cases of stellar lenses with radius $R_\star$. Using the shorthand
\begin{align}
\tilde\theta_1=\sqrt{\frac{2r_{g1}z_1}{z_{12}(z_{12}+z_1)}},
\end{align}
we can write
\begin{align}
J_0(k\tilde\theta_1 b_2)\simeq\frac{1}{\sqrt{\tfrac{1}{2}\pi k\tilde\theta_1 b_2}}\cos (k\tilde\theta_1 b_2-\tfrac{1}{4}\pi).
\end{align}
Then,
\begin{align}
{\cal A}_{\tt 2GL}(\rho)&{}\simeq\frac{k}{\tilde{z}_2}
\frac{b_2\sqrt{2\pi kr_{g1}}}{\sqrt{\tfrac{1}{2}\pi k\tilde\theta_1 b_2}}J_0\left(k\frac{b_2}{\tilde{z}_2}\rho\right)
\int_0^\infty db_2~
\cos (k\tilde\theta_1 b_2-\tfrac{1}{4}\pi)
\exp\left[ik\left(\frac{b_2^2}{2\tilde{z}_2}-2r_{g2}\ln kb_2\right)\right]\nonumber\\
&{}=\frac{k}{\tilde{z}_2}\frac{b_2\sqrt{2\pi kr_{g1}}}{\sqrt{2\pi k\tilde\theta_1 b_2}}J_0\left(k\frac{b_2}{\tilde{z}_2}\rho\right)\nonumber\\
&{}\times\int_0^\infty db_2~
\Bigg\{
\exp\left(i\left[k\left(\frac{b_2^2}{2\tilde{z}_2}-2r_{g2}\ln kb_2+\tilde\theta_1 b_2\right)-\tfrac{1}{4}\pi\right]\right)\nonumber\\
&\hskip 0.75in{}+\exp\left(i\left[k\left(\frac{b_2^2}{2\tilde{z}_2}-2r_{g2}\ln kb_2-\tilde\theta_1 b_2\right)+\tfrac{1}{4}\pi\right]\right)
\Bigg\}.
\end{align}
The phase terms in this expression are given by
\begin{align}
\Phi_\pm=k\left(\frac{b_2^2}{2\tilde{z}_2}-2r_{g2}\ln kb_2\pm\tilde\theta_1 b_2\right)\mp\tfrac{1}{4}\pi.
\end{align}
Differentiating yields
\begin{align}
\frac{d\Phi_\pm}{db_2}=\frac{k}{\tilde{z}_2}b_2-2kr_{g2}\frac{1}{b_2}\pm k\tilde\theta_1=0,
\end{align}
solved, using $\theta_2=\sqrt{2r_{g2}/\tilde{z}_2}$, by
\begin{align}
b_2=\tfrac{1}{2}\tilde{z}_2\left(\sqrt{\tilde\theta_1^2+4\theta_2^2}\mp\tilde\theta_1\right)\simeq\tilde{z}_2(\theta_2\mp\tfrac{1}{2}\tilde\theta_1),
\end{align}
where the final approximation was obtained noting that $\tilde\theta_1\ll\theta_2$. Substituting these back into the expression for $\Phi$, we obtain
\begin{align}
\Phi_\pm=k\left[\tfrac{1}{2}\tilde{z}_2(\theta_2^2-\tfrac{1}{4}\tilde\theta_1^2)
-2r_{g2}\ln k\tilde{z}_2(\theta_2\mp\tfrac{1}{2}\tilde\theta_1)
\right]\mp\tfrac{1}{4}\pi.
\end{align}

The second derivative of the phase term is given by
\begin{align}
\frac{d^2\Phi_\pm}{db_2^2}=k\left(\frac{1}{\tilde{z}_2}+\frac{2r_{g2}}{b_2^2}\right).
\end{align}
Thus, the stationary phase approximation yields a solution in the form
\begin{align}
{\cal A}_{\tt 2GL}&{}\simeq\frac{k}{\tilde{z}_2}
\frac{\sqrt{2\pi kr_{g1}}}{\sqrt{2\pi k\tilde\theta_1 \tilde{z}_2(\theta_2\mp\tfrac{1}{2}\tilde\theta_1)}}J_0\left(k(\theta_2\mp\tfrac{1}{2}\tilde\theta_1)\rho\right)\nonumber\\
&~~~{}\times\sqrt{\frac{2\pi}{k\left(\displaystyle\frac{1}{\tilde{z}_2}+\frac{2r_{g2}}{\tilde{z}_2^2(\theta_2\mp\tfrac{1}{2}\tilde\theta_1)^2}\right)}}\left[\tilde{z}_2(\theta_2-\tfrac{1}{2}\tilde\theta_1)e^{i(\Phi_++\tfrac{1}{4}\pi)}+\tilde{z}_2(\theta_2+\tfrac{1}{2}\tilde\theta_1)e^{i(\Phi_-+\tfrac{1}{4}\pi)}\right]\nonumber\\
&{}\simeq
\sqrt{2\pi kr_{g1}}
J_0\left(k\theta_2\rho\right)\sqrt{\frac{\theta_2}{2\tilde\theta_1}}
\left[e^{i(\Phi_++\tfrac{1}{4}\pi)}+e^{i(\Phi_-+\tfrac{1}{4}\pi)}\right].
\end{align}
As before, the corresponding PSF is given as the square of this (now complex) amplitude, that is to say,
\begin{align}
{\tt PSF}_{\tt 2GL}\simeq
2\pi kr_{g1}
J_0^2\left(k\theta_2\rho\right)\frac{\theta_2}{\tilde\theta_1},
\end{align}
where we used the fact that $\Phi_+-\Phi_-\simeq -\tfrac{1}{2}\pi$, again ignoring minor deviations due to the (very) slowly changing logarithmic (Shapiro) term. Substituting, in the limit $z_{12}\gg z_1$, $\tilde\theta_1\simeq \sqrt{2r_{g1}z_1}/z_{12}=b_1/z_{12}$, we obtain
\begin{align}
{\tt PSF}_{\tt 2GL}\simeq
2\pi kr_{g1}
J_0^2\left(k\theta_2\rho\right)\theta_2\frac{z_{12}}{b_1}.
\end{align}
Again, for large arguments, $\left<J_0^2(x)\right>\simeq 1/\pi x$, allowing us to average this PSF as
\begin{align}
\left<{\tt PSF}_{\tt 2GL}\right>=
\frac{2r_{g1}z_{12}}{b_1\rho}=\frac{\theta_1 z_{12}}{\rho},
\end{align}
which is identical to the averaged PSF of the single lens in the $z_1\ll z$ limit.

Thus once again we ascertained, this time using the wave theory of light, that the two-lens ``bridge'' collects no additional light from the source in comparison with a single lens at the source, and does not increase light amplification.

\begin{figure*}[t]
\begin{center}
\includegraphics[scale=0.75]{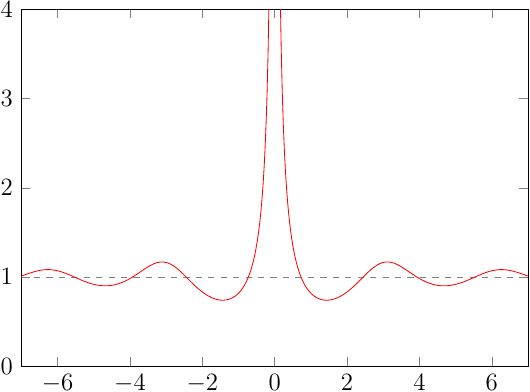}
\end{center}
\caption{\label{fig:PSF1avg}Ratio of the averaged PSF approximation of a single lens vs. its true PSF, both integrated over a circular aperture (\ref{eq:PSFcompare}). Horizontal axis is the argument of the Bessel $J_0$ function in (\ref{eq:PSF1}), proportional to the aperture radius. The approximation works well for large apertures, but diverges in the small-aperture limit.}
\end{figure*}

There is, however, a subtlety that is revealed only through this wave theoretical analysis. The diffraction pattern in the image plane of a single lens in the $z_1\ll z$ limit can be quite large \cite{Turyshev2023}. If the light field of the single lens is sampled by a small aperture telescope, averaging the PSF is no longer justified. The average (\ref{eq:PSF1avg}) diverges near the origin, overestimating the gain. Fig.~\ref{fig:PSF1avg} presents the ratio of the averaged vs. the ``true'' PSF, integrated over a circular aperture $d_A=d_B=d$, that is,
\begin{align}
&\frac
{2\pi \displaystyle\int_0^{d_A/2}\rho~d\rho~\left<{\tt PSF}_{\tt 1GL}\right>}
{2\pi \displaystyle\int_0^{d_B/2}\rho~d\rho~{\tt PSF}_{\tt 1GL}}
\nonumber\\
&\hskip 0.5in{}=\frac{4d_A\sqrt{\dfrac{z(z+z_1)}{2r_gz_1}}}{\pi k d_B^2\left[J_0\left(\tfrac{1}{2}kd_B\sqrt{\dfrac{2r_gz_1}{z(z+z_1)}}\right)^2+J_1\left(\tfrac{1}{2}kd_B\sqrt{\dfrac{2r_gz_1}{z(z+z_1)}}\right)^2\right]}.
\label{eq:PSFcompare}
\end{align}
As we can see, the averaged PSF approximation matches the true PSF well except for the central region of this plot, corresponding to small apertures. The average begins to overshoot the true PSF when $k\sqrt{2r_gz_1/z(z+z_1)}(d/2)\lesssim 0.7$. By way of example, when $\lambda=2\pi/k=600~$nm, $z=10$~ly and $z_1=650$~AU with a $1~M_\odot$ lens,
having a $d_A=1$~m aperture in the focal region of the ``bridge'' delivers the same gain as a $d_B\simeq 4$~m aperture looking directly at the first lens (no ``bridge'').
The effect scales with distance, wavelength, and the inverse of the aperture but remains modest; even at $z=10$~pc, $\lambda=1~\mu$m, an aperture of $d_B\simeq 9$~m is sufficient to obtain the same gain as having a second gravitational lens in a ``bridge'' configuration with a $d_A=1$~m aperture at its focal region.

\section{Ray tracing}
\label{sec:raytrace}

As mentioned previously, ray tracing, a standard family of algorithms used extensively in computer graphics, while useful to study the qualitative features of an optical system, cannot reliably recover its light amplification. Quite simply, standard ray tracing can help determine the light source from which light arrives at the observing location from a given direction, but not the intensity of this light, which may be affected by diffraction en route, including diffraction by the gravitational field.

Nonetheless, we endeavored to build a ray tracing model to help us visualize multiple non-coplanar lenses. We were driven, in part, by the desire to confirm the emergence of a double Einstein ring in the simplest case (source, lenses, observer being colinear) but soon we recognized that even a small number of non-coplanar lenses can yield unanticipated complexity in the resulting projection.

\begin{figure*}[t]
\begin{center}
\includegraphics{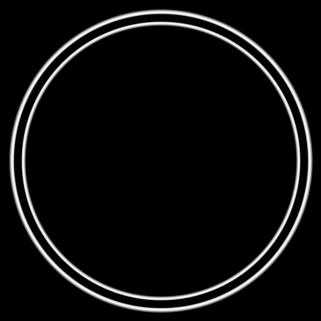}
\end{center}
\caption{\label{fig:twolens}The fully axisymmetric two-lens bridge yields two concentric Einstein rings of approximately equal brightness, as depicted in this simulation. Note that the geometric parameters of the system were greatly exaggerated for visibility; in an actual astrophysical two-lens system, the two Einstein rings would be very close to each other, only appearing as distinct rings at extreme angular resolution. (NB: The lensing objects are assumed to be dark and opaque, hence no direct line-of-sight image of the source appears at the image center.}
\end{figure*}

Needless to say, visual modeling an actual astrophysical two-lens bridge is not feasible: the separation of the two concentring Einstein rings is far too small to be visible at any reasonable resolution. However, it is possible to model exaggerated cases that visually demonstrate how the two-lens bridge works. Just such a case is depicted in Fig.~\ref{fig:twolens}.

\begin{figure*}[t]
\begin{center}
\includegraphics{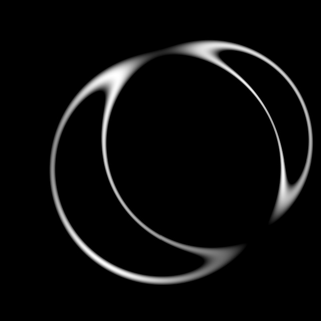}
\end{center}
\caption{\label{fig:twolens-displaced}The two-lens bridge with both lenses displaced by a random amount.}
\end{figure*}

Although ray tracing cannot really inform us about light amplification, it can, on the other hand, be used to model deviations from axial symmetry with ease. A deviation of this type is depicted in Fig.~\ref{fig:twolens-displaced} with both lenses displaced relative to the optical axis. The result in this case appears to be in the form of two intersecting, partial Einstein rings, each consisting of two unequal segments.

\begin{figure*}[t]
\begin{center}
\includegraphics{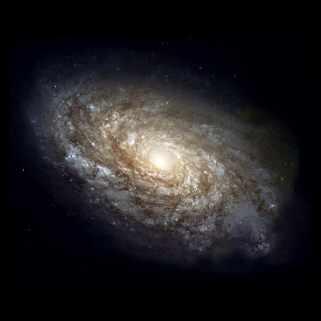}~~~\includegraphics{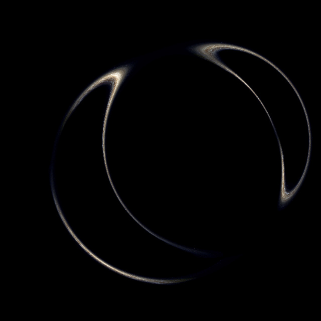}
\end{center}
\caption{\label{fig:twolens-displaced-ngc4414}An actual astronomical object, the galaxy NGC-4414 (left) and its simulated view (right), a seen through the same displaced two-lens system shown in Fig.~\ref{fig:twolens-displaced}. Though it is by no means evident, the result shows four distinct, partially overlapping images of the galaxy.}
\end{figure*}

We are not restricted to imaging compact homogeneous objects. Fig.~\ref{fig:twolens-displaced-ngc4414} shows the image of a galaxy, NGC-4414, displaced through the same lens shown in Fig.~\ref{fig:twolens-displaced}. The four segments of the two overlapping partial Einstein rings each contain a complete, distorted image of the original galaxy.

\begin{figure*}[t]
\begin{center}
\includegraphics{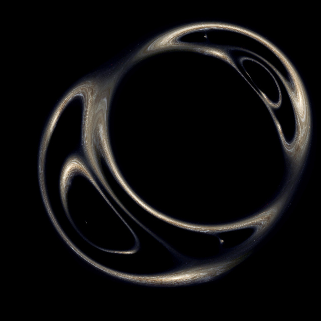}
\end{center}
\caption{\label{fig:quadlens-ngc4414}Simulated view of NGC-4414, as seen through a combination of four lenses, each slightly displaced from the optical axis. Though far from obvious, this image consists of sixteen distinct, distorted views of NGC-4414.}
\end{figure*}

And of course we can easily include additional lenses, yielding images of surprising complexity. Fig.~\ref{fig:quadlens-ngc4414} depicts the same galaxy, NGC-4414, its light distorted by a combination of four non-coplanar, non-colinear lenses.

Since real astrophysical lenses often have similar, complex structure, characterized by a three-dimensional mass distribution, the question naturally arises: Is it possible to reconstruct a view of the original galaxy without distortion from these views that contain multiple images of the same object? The answer, unfortunately, is that in the general case it is likely not possible, because a degeneracy exists between the three-dimensional mass distribution of the lens and the properties of the lensed object. We explored this topic in \cite{SGL2021g}, where we found that at best, we can only recover partial information about the mass distribution of the lens, effectively a projection of the lens onto the two-dimensional surface of the lens plane. Therefore, unless we have additional information available either about the lens or about the target (or both), unambiguous reconstruction is not possible.

\section{Discussion}
\label{sec:disc}

In this paper, we analyzed combinations of multiple gravitational lenses using a variety of methods.

First, we extended a previously developed formalism for the geometric optics analysis of a gravitational lens in the paraxial, thin-lens, weak gravitational field approximation, demonstrating that, in particular, a two-lens system (the so called gravitational lens ``bridge'') delivers no advantages, no additional signal amplification over the amplification offered by a single lens near the source.

Next, we re-derived this same result using a related but different technique, an analytic variation of photon mapping, estimating the proportion of light rays that would intercept a given telescope aperture after going through either a single lens or two lenses. Again, we confirmed that the second lens offers no additional light amplification.

Is it possible that the necessary approximations that form the foundation of these geometric methods represent an oversimplification, yielding a false result? To investigate this possibility, we endeavored to analyze the two-lens ``bridge'' using our previously established wave theoretical formalism. Again, we recovered the same result: the two-lens ``bridge'' offers no additional signal amplification when we compare averaged point-spread functions.
Nonetheless, the wave theory reveals that a modest gain may be present when a small-aperture instrument is used for observation. In this case, the averaged PSF overestimates the gain of the single-lens system, so compared to the ``true'' single-lens PSF, the two-lens ``bridge'' appears to offer a minor advantage.


Finally, we moved on to yet another geometric technique, commonly used also in computer graphics, ray tracing. While ray tracing cannot be used to estimate light amplification reliably, it can help us recover qualitative features of a model, including scenarios with broken symmetries that are very difficult to deal with using analytical tools. Using ray tracing, we were able to recover the visual structure of the image produced by a two-lens system (two concentric Einstein rings of approximately equal magnitude), further confirming our result. We were also able to use our recently developed software to generate images representing real astrophysical scenarios, including the possibility of a background object seen through two (or more) lenses, producing multiple distorted images of the source. The techniques used here could also, in principle, be used to model lenses that are not gravitational monopoles, but these cases are beyond the scope of the present study.

The software used to generate the images in this paper has been released as open source and can he downloaded from \url{https://github.com/vttoth/BRIDGE}.

\section*{Acknowledgments}

VTT thanks Slava Turyshev for discussions and acknowledges the generous support of David H. Silver, Plamen Vasilev and other Patreon patrons.

\bibliography{SGLGEOM-newbridge}

\appendix

\section{Ray tracing software implementation}

\begin{figure*}[t]
\includegraphics[scale=1.5]{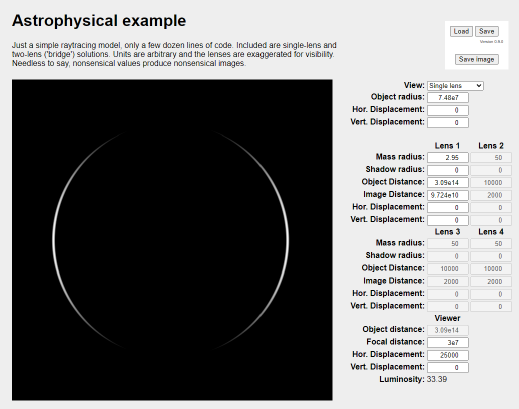}
\caption{\label{fig:astro-object-ui}Gravitational lens ray tracing software implementation}
\end{figure*}

Images presented in this paper were created using software that can be used to visualize images produced by up to four non-coplanar monopole gravitational lenses.

The software was developed using \textsc{JavaScript}, with a simple graphical user interface that can be used through a Web browser (Fig.~\ref{fig:astro-object-ui}). Though initially designed using arbitrary units, the software has the requisite parameter range and accuracy to depict realistic lensing. For instance, the example shown in Fig.~\ref{fig:astro-object-ui} shows lensing by a 1 solar mass lens, of an object, an orb with a radius of 75 million kilometers, with an observing location that is 25,000~km displaced from the optical axis.

The software allows for the saving of parameter sets in the form of \textsc{JSON} objects, which can be reloaded. Images can also be saved. By adjusting parameter values one frame at a time, and saving the corresponding generating images, it is possible to create short animations ``by hand''.

The software also accepts URL parameters, specifically the parameter {\tt img={\em filename}}, which makes it possible to replace the generated white orb with an image chosen by the user. This is how earlier images depicting NGC-4414 were produced for this paper.

Finally, the software displays a ``luminosity'' value, which corresponds to the ratio of image brightness to the most recently displayed ``plain view'' (no lens present) image. This value should be considered as being of an advisory nature only, as ray tracing does not adequately account for light amplification.

The software is available in source form for download at \url{https://github.com/vttoth/BRIDGE}.

\end{document}